\documentclass[aps,prb,twocolumn]{revtex4}   
\usepackage{amsmath}    
\usepackage{graphicx}   

\newcommand{\invcm}{cm$^{-1}$}
\begin{document}

\title{Surface-enhanced Raman scattering of SnO$_{2}$ bulk material and colloidal solutions}
\author{Enza Fazio}
 \affiliation{Dipartimento di Fisica della Materia e Ingegneria Elettronica, Universit\`{a} degli Studi di Messina\\
Viale F. Stagno d\`{}Alcontres  31, 98166 Messina, Italy.
}
\author{Fortunato Neri}
\affiliation{Dipartimento di Fisica della Materia e Ingegneria Elettronica, Universit\`{a} degli Studi di Messina\\
Viale F. Stagno d\`{}Alcontres  31, 98166 Messina, Italy.
}
\author{Salvatore Savasta}
\affiliation{Dipartimento di Fisica della Materia e Ingegneria Elettronica, Universit\`{a} degli Studi di Messina\\
Viale F. Stagno d\`{}Alcontres  31, 98166 Messina, Italy.
}
\author{Salvatore Spadaro}
\affiliation{Advanced and Nano Materials Research s.r.l. \\
Viale F. Stagno d\`{}Alcontres  31, I-98166, Messina, Italy}
\author{Sebastiano Trusso}
\affiliation{CNR-IPCF Istituto per i Processi Chimico-Fisici\\
Viale F. Stagno d'Alcontres 37, 98158, Messina, Italy}
\email{trusso@me.cnr.it}
\date{\today}

\begin{abstract}
Surface enhanced Raman scattering (SERS) effects on tin dioxide in the form of bulk material, nanostructured thin films and colloidal solutions were investigated. Raman spectra are characterized by the three Raman scattering peaks at 478, 633, and 776 \invcm, assigned to the E$_g$, A$_{1g}$ and B$_{2g}$ modes, typical of rutile SnO$_2$. In presence of the silver nanoparticles, in addition to the enhancement intensity of some of the fundamental tin dioxide rutile Raman features, the appearance of a new Raman scattering peak at about 600 cm$^{-1}$ is observed. This spectral features is observed, in presence of silver nanoparticles, also in other SnO$_2$ based system such as pulsed laser deposited thin films, with different stoichiometry, and in water colloidal solutions. The observed SERS effects are explained in terms of electric-field gradient mechanism that are generated near a metal surface. In particular, the appearance of a peak near 600 \invcm~ is attributed to the Raman activation of the infrared E$_u$ transverse optical (TO) mode.
\end{abstract}

\maketitle

\section{Introduction}

In the last decade there has been an increasing interest in materials that can be used as surface enhanced Raman scattering (SERS) substrate capable of appropriate levels of detection to allow the identification of trace constituents (for example drug compounds recovered from surface contaminations).  The enhancement of the Raman signal is mainly achieved by coupling the vibrational modes of the analyte molecule with the electromagnetic field generated at a metallic nanostructure, usually made of gold or silver, upon excitation with light of appropriate energy \cite{1a,2a,3a,xx}. Moreover, recently, the use of a chemically inert shell coating around gold nanoparticles allows to protect the SERS-active nanostructure, determining a much higher detection sensitivity and an increased field of application \cite{4a}. Nevertheless, along with the signal amplification, other spectral effects are often found, whose nature is stricly connected to the material under SERS analysis.

In this paper we investigate the modifications of the Raman spectral response of SnO$_2$ induced by silver nanoparticles. 
  The high stability, the wide band gap, the high sensitivity and the ability of SnO$_2$ to be integrated onto micro-machined substrates  make it a good candidate for the fabrication of dye-based solar cells, transistors, electrode materials, catalytic or electrochromic devices and also miniaturized, ultrasensitive gas sensors due to its high reactivity with environmental gases \cite{3b,3c,3d,3e,3f}. Since the sensing properties of solid state gas sensors are directly related to the surface area, the reduction of SnO$_2$ grains size leads to a much higher surface-to-bulk ratio, thus enhancing both the properties and performance of the devices. 
In an effort to overcome some of the SnO$_2$ technological performance limitations, a systematic study of its structural properties is crucial. As is well known, position, linewidth and lineshape of the Raman peaks are very sensitive to the local arrangements of structured and inhomogeneous materials \cite{3g}.  Hence, Raman spectroscopy results to be  an effective, simple and non-destructive technique to characterize SnO$_2$ micro and nanostructures.  Anomalous Raman spectra, where some IR allowed vibrations become Raman active, were observed in SnO$_2$ nanowires or SnO$_2$ nanostructures loaded by metal nanoparticles \cite{4b,4c,4d}. These findings were related to the grain size, films roughness and, more generally, to structural disorder but, their origin was not exactly determined. 

In addition to the spectral intensity enhancement, SERS determines the shifting and the broadening of the fundamental vibrational Raman modes as well as the appearance of new features. Generally, the SERS induced occurrence of Raman forbidden peaks has been attributed to the lowering of the symmetry of the investigated molecules due to the formation of bonds at the metallic surface or to the presence of a steep field gradient close to the nanoparticles metal  surfaces \cite{5a,5b,5c}.

Here, we analyze the appearance of new Raman-like lines either in SnO$_2$ bulk material and in nanostructured thin films, only in presence of silver nanoparticles.  Moreover, we show that these effects are evident even in water colloidal solutions of differently sized SnO$_2$ particles.
The electric field gradient mechanism, occurring in the near field region of the metallic nanoparticles, was taken into account to explain the experimental evidences.

\section{Experimental section}
\subsection{Samples preparation and characterization}
SERS experiments were carried out on different kinds of SnO$_2$ material: hot pressed pellet, pulsed laser deposited thin films and water colloidal solutions containing either micrometric or nanometric sized particles.
Pellets were obtained by hot pressing commercial micrometric sized powder purchased from Goodfellow. The pulsed laser deposited (PLD) SnO$_2$ thin films were prepared in a vacuum chamber using a 248 nm KrF laser beam focused at an incidence angle of 45$^o$ on the surface of a hot pressed SnO$_2$ pellets used as target. The process was carried out at a substrate temperature of 470 K with a laser fluence of 1.0 Jcm$^{-2}$ in a controlled high purity oxygen gas background atmosphere.  The two samples studied here were grown at the oxygen pressure of 1.3 Pa and 13.3 Pa. The films thickness was around 100 nm while the oxygen content $x$, estimated by X-ray photoelectron spectroscopy, was 1.8 and 2.2, respectively. Further details on the experimental procedures and films properties were reported elsewhere. \cite{1,2,3}. SnO$_2$ water colloidal solutions were prepared following two different procedures: 1) by irradiating an high purity metallic tin target, immersed in distilled water, with the second harmonic of a Nd:YAG laser (wavelength 532 nm, pulse width 5 ns, repetition rate 10 Hz) at the laser fluence of 1.0 Jcm$^{-2}$, for typical irradiation times of 20 min; 2) dispersing 1.5 mg of SnO$_2$ micrometric sized powder in 6 mL of pure water.
Silver nanoparticle (NP) colloidal solutions were prepared in water ablating an high purity silver target, at the laser fluence of 0.7 Jcm$^{-2}$ for a time of 20 min.
To carry out transmission electron microscopy (TEM) and X-ray diffraction (XRD) measurements, few drops of the SnO$_2$ solutions were casted on carbon coated copper grids and silicon substrates, respectively. TEM images were taken by a transmission electron microscope (model JEOL-JEM 2010) operating at an accelerating voltage of 200 kV, while X-ray diffraction(XRD) patterns were recorded, in the 2$\theta$ range from 20$^o$ to 80$^o$, using a Bruker D8 Advance X-ray diffractometer with the Cu K$_\alpha$ radiation (1.5406$^o$$\AA$). UV-vis optical transmission measurements of the silver NP solutions were carried out by a Perkin-Elmer LAMBDA 2 spectrophotometer in the 190-1000 nm range. The estimated Ag particles size was 20 nm, as determined from the position and the full width at half maximum (FWHM) of the surface plasmon resonance absorption peak and confirmed by the TEM imaging analysis.
\subsection{SERS measurements} 
SERS measurements were performed on the surface of the hot pressed pellet and of the PLD deposited thin films after silver colloidal solutions were air-brush sprayed onto their surfaces. SERS measurements on the colloidal SnO$_2$ solutions were performed by adding 3 ml of the silver colloid to an equivalent volume of the SnO$_2$ solution one. A diluition study was carried out by progressively adding 3, 4, 5 and 6 ml of distilled water to the Ag-SnO$_2$ colloid. Raman scattering measurements were carried out in a backscattering geometry by means of a confocal micro-Raman apparatus focusing the 632.8 nm line of an He-Ne laser on the solid samples surface through the 50x objective of a microscope. Raman spectra from the colloidal solutions were collected using a 10x Mitutoyo infinity-corrected long working distance microscope objective. In both cases backscattered radiation was collected by the same microscope optics, dispersed by a Jobin-Yvon Triax 320 monochromator equipped with a 1800 line/mm holographic grating and detected with a LN$_2$ cooled charged coupled device (CCD) sensor. All the shown spectra are normalized using their integration times.

\section{Results and discussion}
\begin{figure}[t]
\begin{center}
\includegraphics[scale=.75]{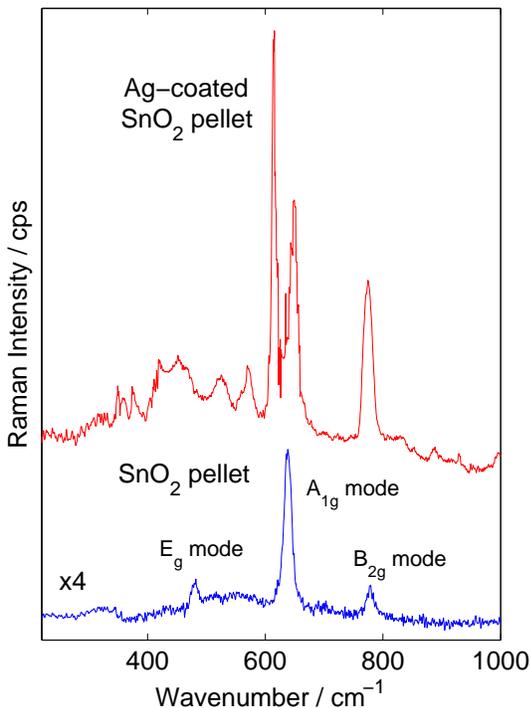}
\caption{\label{fig:1} Raman spectra of SnO$_2$ and Ag-coated SnO$_2$ pellet.}
\end{center}
\end{figure}

Tin dioxide in the rutile form belongs to the symmetry space group D$_{4h}$ \cite{4,5,6,7,8,9}. The 6 units cell atoms give rise to 18 vibrational modes. Two modes are infrared active (the single A$_{2u}$ and the triply degenerate E$_u$), four are Raman active (the three nondegenerate A$_{1g}$, B$_{1g}$, B$_{2g}$ modes and the doubly degenerate E$_{g}$ one) and two others are silent (the A$_{2g}$ and B$_{1u}$ modes). The A$_{1g}$ mode is a vibration along the c axis while the E$_u$ mode involves movements in the xy plane\cite{10,11,12,13,14,15,16}.  Figure \ref{fig:1} shows the Raman spectrum collected on the surface of a pellet obtained by hot pressing SnO$_2$ powder. The spectrum is charaterized by the presence of three peaks located at about 478, 633 and 776 cm$^{-1}$ corresponding to  the E$_g$, A$_{1g}$ and B$_{2g}$ modes in good agreement with those observed in a single-crystal SnO$_2$ or poly-microcrystalline films \cite{17}. Three broader and less intense bands can be observed at 350 \invcm~
 and between 400 and 800 \invcm. They were identified as belonging to surface and acoustic modes, respectively \cite{18,19}.

\begin{figure}[t]
\begin{center}
{\includegraphics[scale=.5]{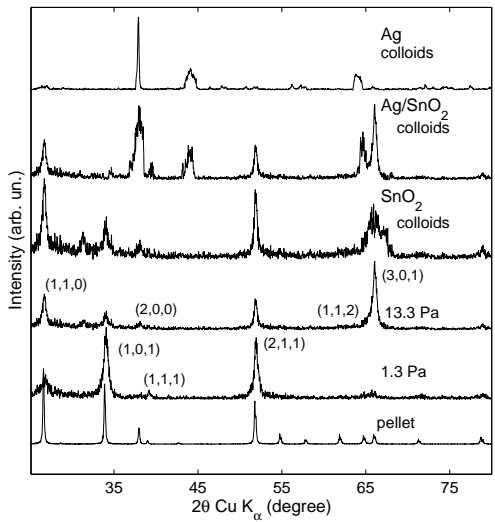}}
\caption{\label{fig:5} XRD spectra of all the investigated samples.}
\end{center}
\end{figure}

In the same figure, the Raman spectrum collected on the hot pressed SnO$_2$ pellet surface after the colloidal Ag air-brush spraying procedure is shown. Beside the above reported three fundamental Raman modes, a new very intense peak at about 615 cm$^{-1}$ is evident. Furthermore, other less intense peaks are present in the low wavenumber portion of the spectrum at about 350, 374, 419, 482 and 524 \invcm. The origin of the intense peak observed at 615 cm$^{-1}$ could be attributed to the presence of the amorphous SnO$_2$ phase, even if, generally, a broad Raman band related to the amorphous phase is located at about 570 cm$^{-1}$ \cite{20}.  Nevertheless, its position, narrow linewidth and intensity, greater than the A$_{1g}$ one, rule out such an attribution. Another interesting effect of the presence of the silver NPs on the samples surface is the enhancement of the B${_{2g}}$ mode related peak whose intensity becomes comparable to the A$_{1g}$ one. Otherwise, the E$_g$ mode peak intensity resulted unaffected being hindered by the appearance of the above mentioned low frequency peaks. Concerning the observed low wavenumber   peaks, their origin is still unclear and debated in literature. In the low-dimensional SnO$_2$ systems their presence was associated to the activation of Raman modes as a consequence of disorder or size effects \cite{4d,20}. Taking into account that the hot pressed pellet was obtained from stoichiometric powders with particles having typical dimensions in the micrometric range, we can safely rule out any size effect. Moreover, it's worth to mention that we are able to detect such bands only in presence of the silver NPs. 
\begin{figure}[t]
\begin{center}
{\includegraphics[scale=0.5]{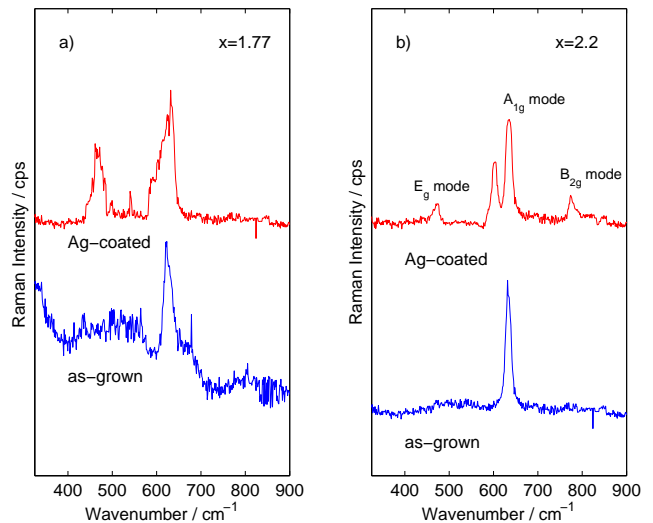}}
\caption{\label{fig:2}Raman spectra of SnO$_x$ and Ag-coated SnO$_x$ PLD grown samples.}
\end{center}
\end{figure}
The same silver NPs deposition procedure was performed on the surface of two SnO$_x$ thin films grown by pulsed laser deposition at different deposition conditions. The films are characterized by different oxygen content $x$, $i.e.$ 1.8 and 2.2 as determined by XPS measurements. The different oxygen content arised from the oxygen partial pressure ($i.e.$ 1.3 and 13.3 Pa) at which the samples were grown. More details about the growth process can be found in Trusso $et$ $al.$\cite{1}. The corresponding Raman spectra are shown in Fig. \ref{fig:2}. In the Raman spectra of the as deposited samples only the A$_{1g}$ peak is easily distinguishable, being its linewidth narrower for the sample with $x$=2.2. A broad and structureless band is also barely visible in 400-600 cm$^{-1}$ region for both the samples. After the silver colloidal solution was sprayed on the films surface, the corresponding Raman spectra resulted very different. In the sub-stoichiometric film the E$_g$ mode become visible while the A$_{1g}$ mode related band widens showing a shoulder on the low wavenumber side (see Fig. \ref{fig:2}a). Concerning the other film, two effects are evident: the appearing of the $E_g$ and $B_{2g}$ mode related bands, together with a peak at about 600 cm$^{-1}$, as already observed in the Raman spectra of the hot pressed pellet. 
It should be noted that, at the surface of the sub-stoichiometric film, the presence of oxygen vacancies can be envisaged. Taking into account that the E$_g$ mode originated by the vibration of two oxygen atoms along the $c$ axis, but in opposite direction, such a mode is more sensitive to oxygen vacancies than the other modes: the broad bump observed at around 500 cm$^{-1}$ for E$_g$ mode proves this point. As reported in a previous work \cite{3}, all the PLD deposited films are composed of nanoparticles exceeding 10 nm in size and essentially with a tetragonal rutile crystalline structure.  From the analysis of SAED patterns, it was shown that the tetragonal unit cell shrinks along both the principal axes when the compositional parameter $x$ gets smaller, $i.e.$ upon increasing the number of octahedrally arranged vacant sites of the oxygen atoms. Hence we are reasonably confident that, for the sub-stoichiometric film, the appearance of the new peak near the A$_{1g}$ mode and the marked enhancement of the E$_g$ one may be due in part to the surface defect states  which easily linked with the A$_g$ ones. On the other hand, for the $x$= 2.2 stoichiometry sample, the well defined and skrinked features around 600 cm$^{-1}$ and the TEM imaging analysis \cite{3} suggest that the observation of  such vibrational mode is not due to the presence of structural defects. 
\begin{figure}[t]
\begin{center}
{\includegraphics[scale=.7]{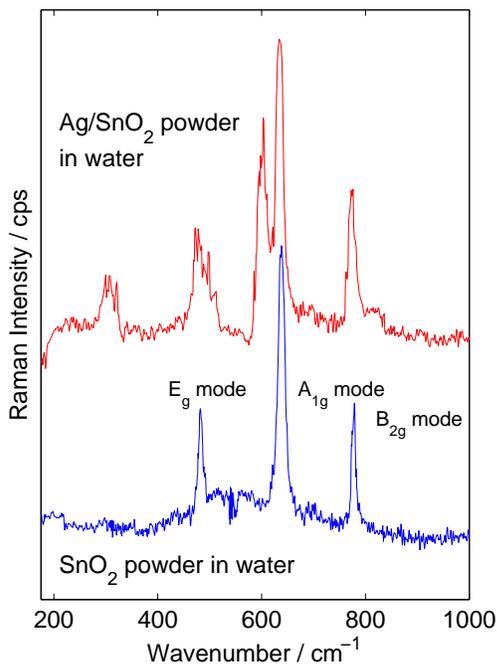}}
\caption{\label{fig:3} Raman spectra of SnO$_2$ and Ag-SnO$_2$ colloidal water solutions.}
\end{center}
\end{figure}
 The investigation was extended to systems consisting of SnO$_2$ colloidal solutions prepared by different methods. A first colloidal solution was prepared simply by dispersing SnO$_2$ micrometric powder in distilled water, while a second colloidal solution was prepared by pulsed laser ablating a pure tin target in water, following the same procedure adopted in the case of the Ag nanoparticles. The SnO$_2$  particles formation is due to reactions between oxygen atoms, originating from dissociation of water molecules in the laser generated plasma reagion, and tin ones coming ejected from the target by the focused laser pulse.  In this latter case the size of the SnO$_2$ colloidal particle resulted in the nanometer range.
 In Fig. \ref{fig:3} the Raman scattering features for water dispersed micrometric SnO$_2$ powder are shown. In the same figure the Raman spectrum acquired after the addition of the silver NPs colloidal solution is reported. The pure micrometric SnO$_2$ powder solution spectrum is almost identical to that acquired on the surface of the pellet (see Fig.\ref{fig:1}). Looking at the Raman spectrum of the silver colloid SnO$_2$ powder mixed solution, the appearance of two peaks at about 300 cm$^{-1}$ and 600 cm$^{-1}$ can be envisaged. Furthermore, the peak centred at about 478 cm$^{-1}$ appears more structured respect to that observed in absence of silver nanoparticles.
\begin{figure}[h]
\begin{center}
{\includegraphics[scale=0.18]{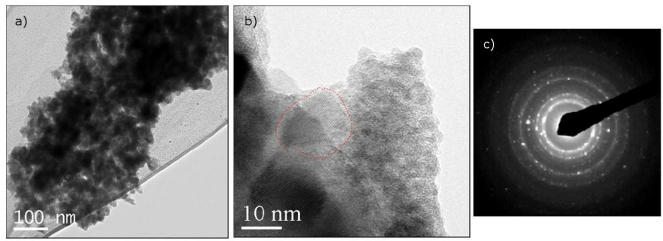}}
\caption{\label{fig:4}TEM images and SAED patterns of the SnO$_2$ colloidal solutions.}
\end{center}
\end{figure}
 A structural characterization of the different SnO$_2$ systems was performed by TEM imaging and selected-area electron diffraction (SAED) measurements. The results are reported in Fig. \ref{fig:4} where TEM images and SAED patterns of dried SnO$_2$ nanoparticles are shown. In the TEM image clusters, 25 nm in size, composed by nearly spherical nanoparticle can be observed. As shown by the HRTEM image, 6-10 nm sized nanoparticles are present; in some cases, they overlap and are connected with two or three neighbors through necks. SAED patterns measurements were carried out to look at the crystallographic structure of the colloids. SAED patterns show the (1,1,0), (1,0,1) and (2,1,1) rings typical of the tetragonal rutile SnO$_2$ structure. The widespread halo cannot be associated with reasonable certainty to the SnO$_2$ amorphous phase since the carbon membranes, onto which the colloidal solution was deposited, could contribute to the total homogeneous and isotropic diffraction. Looking for a confirm concerning the crystallographic structure of the samples, XRD measurements were carried out. In Fig. \ref{fig:5} the XRD spectrum of the SnO$_2$ colloids and of the samples deposited in oxygen atmosphere are shown for comparison. The normal tetragonal rutile structure with major reflections along the (2,1,1), (1,0,1) and (3,0,1) planes occurs for all the samples. For the SnO$_2$ colloids, the reflections (2,1,1) and (1,1,0) become predominant. The same analysis, carried out for the Ag-SnO$_2$ mixed colloids, shows, in addition, the typical silver phase features corresponding to the (1,1,1), (0,0,2) and (2,2,0) reflections referred to the peaks occurring around 37.95$^o$$\AA$, 44.15$^o$$\AA$ and 64.30$^o$$\AA$. Then, we have no evidence that the chemistry of the tin oxide material is affected by the presence of the silver nanoparticles.

\begin{figure}[t]
\begin{center}
{\includegraphics[scale=.5]{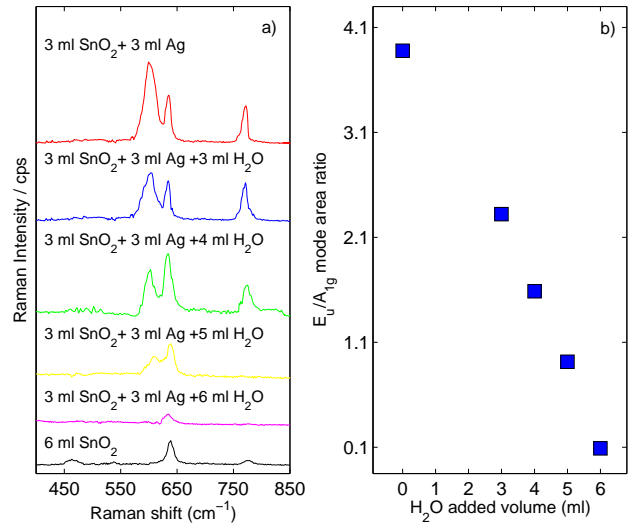}}
\caption{\label{fig:6}  a) Raman spectra of Ag-SnO$_2$ colloidal solutions; b) E$_u$/A$_{1g}$ mode area ratio $vs$ H$_{2}$O added volume.}
\end{center}
\end{figure}

The Raman spectrum of the nanostructured SnO$_2$ colloidal solution is shown in Fig. \ref{fig:6}a. The main features is the well defined peak centered around 630 cm$^{-1}$ while the contributions related to the E$_g$ and B$_{2g}$ are barely visible. By adding a silver nanoparticles solution, in addition to the enhancement intensity of some of the fundamental tin dioxide rutile Raman features, a new well-defined peak towards 600 cm$^{-1}$ appears and grows in intensity at expense of the A$_{1g}$ mode contribution.

 In order to rule out any chemical effect between Ag atoms and the surface of SnO$_2$ particles, we diluted the Ag-SnO$_2$ solution by adding progressively few ml of water. The resulting spectra are reported in Fig. \ref{fig:6}a. If the activation of the peak observed around 600 cm$^{-1}$ should have a chemical origin (with the formation, for example, of Ag-SnO$_2$ complexes), it should still be present after dilution. On the other side, if the effect has an electromagnetic origin: $i.e.$ due to the strong electric field enhancement, and/or to its gradient, around the metallic nanoparticles induced by localized surface plasmon resonances, the behaviour should be different.  In this latter case, the activation of the normally forbidden Raman mode is effective only when the SnO$_2$ nanoparticles are, at most, only a few nanometers separated by the silver ones. Hence in a colloidal solution the effect should be proportional to the number of effective collisions between the metallic and the SnO$_2$ nanoparticles, during the spectra acquisition time, collisions which are inversely proportional to the added water volume. Indeed Fig. \ref{fig:6}a shows that, upon increasing dilution, the peak intensity of the normally forbidden Raman mode decreases. Particularly, increasing dilution the spectra in Fig. \ref{fig:6}a tend towards the ones in the absence of metallic nanoparticles, displaying a fully reversible behavior, so ruling out the presence of chemical processes activating the Raman peak. In addition, Fig. \ref{fig:6}b shows that the ratio between the areas of the E$_u$ and A$_{1g}$ modes actually decreases almost linearly with the H$_2$O added volume. Thus, the reversibility of the process versus the H$_2$O dilution (Fig. \ref{fig:6}b) indicates that no formation of Ag-SnO$_2$ complexes occurs, in good agreement with the XRD data.
The absence of significant SERS induced frequency shifts in the Raman spectra, the reversibility of the effect upon dilution and the XRD results, suggest a negligible chemical interaction between the compound and the silver nanoparticles. On the overall, this spectral effect is independent on the SnO$_2$ size (bulk or nanostructure) or on the surrounding matrix (i.e. solid or water solution) in which the oxide is found. The observation of these anomalous Raman spectra depends only on the presence of the Ag nanoparticles. Thus, the appearance of the normally Raman forbidden mode around 600 \invcm can, more likely, be attributed to the electromagnetic effects induced by the silver nanoparticles. 
In the last years it has been shown that the appearance of normally forbidden vibration modes can be explained by the presence of an electric field gradient in the near field region of metallic nanoparticles.
In general, the SERS induced occurrence of Raman forbidden peaks has been attributed to the lowering of the symmetry of the investigated molecules due to the formation of bonds at the metallic surface or to the presence of a steep field gradient close to the nanoparticles metal surfaces  \cite{21,22}. 

When the electric field varies over the length of the SnO$_2$ bond, the Raman signal can depend upon the polarizability times the field gradient in addition to the usual dependence on the polarizability gradient. The strong field gradient generated at the metallic surface, as already observed in the SERS spectrum of benzene\cite{21,22}, shifts the potential energy of the induced dipole in an asymmetric manner, leading to a coupling with the applied field that lacks a center of symmetry. On the overall, the loss of a center of symmetry eliminates the requirements of the mutual exclusion rule, which dictates that modes can only be either Raman or infrared active. Thus, modes that would normally appear only in the infrared spectrum can appear in the SERS spectrum \cite{23,24,25}. A similar mechanism can occur when the SnO$_2$ colloidal nanoparticles are at most only a few nanometers separated by the silver ones. The strong electric-field gradient near the silver nanoparticles permits a different coupling mechanism between the optical electric field and the vibration. The selection rules for this process differ markedly from the usual Raman selection rules, and the prefactors favour Raman-like observation of strong infrared (not normally Raman active) vibrations. 
Transitions in vibration levels due to coupling with a radiation field are described by the perturbation Hamiltonian $H= \mu E$, where $\mu$ is the dipole moment and E is the electric field.  The first order expansion of the dipole moment $\mu$ in the coordinate of vibration $Q$ is given by the following expression:\cite{26,27}
\begin{equation}
\mu_a=\left[\left(\frac{d\mu_{\alpha}^p}{dQ}\right)+\left(\frac{d\alpha_{ab}}{dQ}\right)E_b+\alpha_{ab}\left(\frac{dE_b}{dQ}\right)\right]Q+\ldots
\label{eq1}
\end{equation}
where the three terms yield the infrared absorption, the Raman and the gradient field Raman (GFR) contributions, respectively. Compared to the classical treatment, the new GFR term takes into account that the electric field $E$ cannot be removed from the derivative term since $E$, varying very rapidly near the metal surface, depends on the coordinate $Q$. In particular, the ratio of the GFR to the Raman terms is approximated as $\alpha/a$, $a$ being a molecular dimension. The jellium approximation of Feibelman\cite{28} indicates that the electric field in proximity of a metallic surface varies by nearly its full amplitude over 0.2 nm. 
The GFR/Raman terms ratio is expected to be $\approx 1$. In our case, for the colloidal solutions, thanks to Brownian motion, silver nanoparticles collide with SnO$_2$ ones. Thus, during the spectra acquisition time, a number of  Ag and SnO$_2$ nanoparticles, depending on the density, spend some time in close proximity. 
An additional GFR-like contribution may arise from the quadrupole Raman term\cite{27}. Since in the equation \ref{eq1} the polarizability multiplies the electric field gradient term, the GFR effect will be proportional to the polarizability and, therefore, it is even more evident in molecules with larger polarizability, such as those characterized by ionic bonds. Among these materials, SnO$_2$ has a fairly high degree of ionic bonds\cite{29}. This intrinsic characteristic of the material is another element that leads us to believe that is reasonable to adopt the electric field gradient mechanism to explain the observed SERS effects.

\section{Conclusions}
Significant surface-enhanced Raman scattering (SERS) effects from SnO$_2$ in the form of bulk material, nanostructured thin films and colloidal solutions have been reported for the first time, at least to our knowledge. In presence of silver nanoparticles, beside the intensity enhancement of some of the fundamental SnO$_2$ rutile Raman features, the appearance of a new Raman-like line at about 600 cm$^{-1}$ is observed. It was found that the appareance of the new Raman peak depends only on the proximity of the Ag nanoparticles to the SnO$_2$ material, indipendently on its grain size and on the surrounding matrix in which the oxide is found, whether in solid phase or in solution. Such a Raman feature was attributed to the infrared active $E_u$ TO mode, normally Raman inactive. It is suggested that it becomes Raman active as a consequence of the gradient field Raman mechanism.  In our opinion the investigations here proposed are interesting for the potential applications of SnO$_2$ in various technological fields.  Furthermore, the investigations will deserve an extension to other systems in order to check for the limits within the GFR picture can be used to explain the SERS induced spectral modifications.


%
%
%
%
%
%
%
\end{document}